\documentclass[12pt,a4paper]{report}

\usepackage[utf8]{inputenc}
\usepackage[T1]{fontenc}
\usepackage[english]{babel}  
\usepackage[usenames]{color}
\usepackage{xcolor}
\usepackage{pdfpages}
\usepackage{amssymb} 
\usepackage{amsthm} 
\usepackage{amsmath} 
\usepackage{multicol} 
\usepackage{esint}
\usepackage[a4paper]{geometry} 
\usepackage{graphicx}
\usepackage{array} 
\usepackage{paralist} 
\usepackage{verbatim} 
\usepackage{subfigure} 
\usepackage{hyperref}
\usepackage{frcursive}
\usepackage{lmodern}
\usepackage{caption}
\usepackage{mathtools}
\usepackage{csquotes}
\usepackage{setspace}
\usepackage{stmaryrd}
\usepackage[toc,page]{appendix}

\makeatletter 
\renewcommand{\maketitle} 
{ \begingroup \vskip 10pt \begin{center} \large {\bf \@title}
	\vskip 10pt \large \@author \hskip 20pt \@date \end{center}
  \vskip 10pt \endgroup \setcounter{footnote}{0} }
\makeatother 

\newcommand{\abs}[1]{\left| #1 \right|} 
\newcommand{\avg}[1]{\left< #1 \right>} 
 
\let\baraccent=\= 
\renewcommand{\=}[1]{\stackrel{#1}{=}} 

\theoremstyle{definition}

\theoremstyle{remark}

\begin{document}

\title{\fbox{%
   \begin{minipage}{\textwidth}
      \begin{center}{A comparison of $g^{(1)}(\tau)$, $g^{(3/2)}(\tau)$, and $g^{(2)}(\tau)$, for radiation from harmonic oscillators in Brownian motion with coherent background}
      \end{center}
   \end{minipage}%
}}

\author{A. Siciak,  L.A.Orozco, M. Fouch{\'e}, W. Guerin, R. Kaiser}
\date{June 15, 2020}
\maketitle

\begin{abstract}
We compare the field-field $g^{(1)}(\tau)$, intensity-field $g^{(3/2)}(\tau)$, and intensity-intensity $g^{(2)}(\tau)$ correlation functions for models that are of relevance in astrophysics.  We obtain expressions for the general case of a chaotic radiation, where the amplitude is Rician based on a model with an ensemble of harmonic oscillators in Brownian motion. We obtain the signal to noise ratios for two methods of measurement. The intensity-field correlation function signal to noise ratio scales with the first power of $\abs{g^{(1)}(\tau)}$. This is in contrast with the well-established result of $g^{(2)}(\tau)$ which goes as the square of $\abs{g^{(1)}(\tau)}$.
\end{abstract}

\newpage
\tableofcontents

%
\newpage

\section{Introduction}\label{Introduction}
Correlation functions appear in the study  of Physics from the microscopic scale to the galactic scale, and have played fundamental roles in statistical physics at equilibrium, and out of equilibrium 
. The intimate relation between correlations and spectra is now one of the most solid foundations in many areas of science. To name only one example, the 2019 Nobel prize of J. Pebbles, for his understanding of the large-scale anisotropy of the Microwave Black Body spectrum, and for the mass fluctuations at large scale 
\cite{peebles1982large}. 

Correlation functions of the electromagnetic field have been applied to classical and quantum fields, their relation to fluctuations can not be over stressed, and their study illuminates the boundary between classical and quantum fields \cite{mandel95}. Intensity-intensity correlation functions started in optics and astrophysics with the pioneer work of Hanbury Brown and Twiss \cite{hanbury56ast,hanbury57,hanbury58}. Measurements give information about the radiative source, such as its size and its spectra, but also about the radiation processes themselves \cite{Dussaux,Lemieux99}. The work of Handbury Brown and Twiss triggered the birth of quantum optics \cite{hanbury56op}, consolidated by the foundational work of R. J. Glauber with the quantum theory of coherence \cite{
glauber63b}. During the last decade there has been strong interest to revive the technique of Hanbury Brown and Twiss in astronomy \cite{Foellmi:2009,Dravins:2016,Zampieri:2016,Pilyavsky:2017,Tan:2016,Guerin:2017,Tan:2017,Guerin:2018,Rivet:2018,Wang:2018,Matthews:2018,Matthews:2019,rivet2020}.
The intensity-field correlation developed from ideas that started to appear some thirty years ago {\it{e.g.}} \cite{yurke87,vogel91,vyas00}. Its formal introduction as conditional homodyning by Carmichael {\it et al.} \cite{carmichael00}, including its relationship with the spectrum of squeezing, triggered experimental studies in cavity QED\cite{foster00} and in single atom resonance fluorescence \cite{gerber09}. Among the theoretical advances followed by Carmichael's work has been  the discovery of its link to weak value measurements \cite{wiseman02,kofman2010}. It would seem that a correlation of three fields would not give any further information, but strictly speaking, there are three fields from the source and one from a local oscillator that cancels out in the normalized definition, giving access to such intensity-field correlation that probes the field quadrature fluctuations. Recent investigations now include the study of Raman scattering of molecules on surfaces \cite{santos2019}.  The paper by Xu {\it {et al.}} \cite{xu2015}  summarizes the motivation for the development of the intensity-field correlation function as a tool to study conditional dynamics in quantum optics. It presents an overview of the connection of the correlation functions, conditional measurements, and photoelectric detection.

This theoretical work
compares the field-field $g^{(1)}(\tau)$, intensity-field $g^{(3/2)}(\tau)$, and intensity-intensity $g^{(2)}(\tau)$ correlation functions, for a simple  system that could meet radiative properties of the stellar emission lines: an ensemble of harmonic oscillators in Brownian motion with a coherent background. The resulting field is chaotic as defined by Glauber \cite{glauber2007}. We add a coherent background for two reasons. First, in astrophysics if coherent radiation exists, it could appear with the blackbody radiation (zero-mean chaotic field) \cite{Letokhov:2009,Tan:2017}. Then, in order to appropriately measure the intensity-field correlation on a blackbody radiation, it is necessary to add a coherent offset to it. The result is chaotic radiation, where the amplitude has a Rician distribution (and not the Rayleigh distribution of the blackbody radiation). We support our results with a simulation based on an elementary model from kinetic theory: the so-called one-dimensional gas at thermal equilibrium, that permits exact calculations and  Monte-Carlo simulations.

Then, we derive the signal to noise ratios and compare them for two different methods of measurement, each relevant in one particular intensity regime. Based on the recent experimental results of our group \cite{Guerin:2017} we take into account only statistical ``shot noise''. 
The study shows the advantages of considering one correlation function or another, either in terms of data collection time, or in terms of physical information that one obtains. At first sight, one could argue that only the first order correlation function $g^{(1)}(\tau)$ really matters. Indeed, the correlation functions of the Gaussian processes are simply linked among them, it is Isserlis' theorem \cite{Isserlis1918} (classical version of Wick's theorem). The knowledge of the first order correlation function is enough in principle. However, we shall see that the physical information appears with slightly different form in each of the correlation functions and that the technical challenges for measuring them differ.

The paper is organized as follows. We first present our theoretical framework in section \ref{Theory}, then in section \ref{chaoticstatesection} we derive these correlation functions, in section \ref{sectionsimu} we present the kinetic model and our simulation, confirming the analytical results. In section \ref{sectionSNR}, we derive the signal to noise ratios, taking into account only the statistical ``shot-noise'', and compare them.
\newpage
\section{Theoretical Framework}\label{Theory}
\subsection{Our classical picture}
We consider an ensemble of massive particles (harmonic oscillators hereafter) in (non-relativistic) Brownian motion, that each emit the same electric field with a fixed polarisation, and some vector modes function satisfying the wave equation with a given gauge choice, with the usual boundary and transversality conditions. We assume that the harmonic oscillators are too few in front of the surrounding thermalized particles to interact among them, they are independent from each other. We assume that the field is disconnected from the particles, such that the Fourier coefficients in the analytic representation of the field fully describe the field \cite{glauber2007}. In all this paper, we consider a single vector component and only one mode. We set an inertial observer at a fixed point in space, we adopt an Eulerian description of the field.
\\In its stable state, the Brownian motion induces, for any harmonic oscillator $\ell$, independent fluctuations in frequency $\left\{\delta\omega^\ell_t\right\}$ around a mean value $\avg{\omega}$, and fluctuations in phases $\left\{\varphi^\ell_t\right\}$ via, respectively, the Doppler-Fizeau mechanism and the elastic collisions mechanism. Then, the field emitted by an harmonic oscillator $\ell$ is described as a continuous time stable Markovian process
\begin{equation}
    \mathcal{E}_\ell(t)= A_\ell(t) \exp{(-i\avg{\omega} t)},
    \label{TheField}
\end{equation}
where the complex Amplitude is:
\begin{equation}
    A_\ell(t)=E_0\exp{(i\varphi^\ell_t)}\exp{(-i\delta \omega^\ell_t t)}.
    \label{TheCAmpl}
\end{equation}
The fluctuations $\left\{\delta\omega^\ell_t\right\}$ are normally distributed because of the Doppler-Fizeau mechanism, and the phases $\left\{\varphi^\ell_t\right\}$ are uniformly distributed because of the elastic collisions mechanism. 
The field resulting from the superposition of $N\gg1$ fields (\ref{TheField}),
\begin{equation}
    \mathcal{E}_{\text{Ra}}(t)=\sum_{\ell=1}^{N\gg1} \mathcal{E}_\ell(t)= A_{\text{Ra}}(t) \exp{(-i\avg{\omega} t)}
    \label{TheFieldRA}
\end{equation}
has a Gaussian complex amplitude $A_{\text{Ra}}(t)$, and so is a chaotic field as defined by Glauber \cite{glauber1965}. The modulus of $A_{\text{Ra}}(t)$ follows a Rayleigh probability law at each time.
\\The analytic representation (\ref{TheField}) allows the classical correspondence \cite{mandel95} of the quantum field-correlation functions defined by Glauber \cite{
glauber63b}
. For a field $\mathcal{E}(t)$, the classical correspondence of the intensity operator is $I(t) \coloneqq \mathcal{E}^*(t)\mathcal{E}(t)$. It is proportional to the energy resident in the radiation. 

\subsection{Correlation functions definitions}
The three first non-normalized correlation functions are defined (when they exist) in the steady state of a field $\mathcal{E}(t)$. Within our Eulerian description they are:
\begin{eqnarray}
    G^{(1)}(\tau) \coloneqq \avg{\mathcal{E}^*(t)\mathcal{E}(t+\tau)}\label{defG1}\\
    G^{(3/2)}(\tau)\coloneqq\left<\mathcal{E}^*(t)\mathcal{E}(t)\mathcal{E}^*(t+\tau)\mathcal{E}_{lo}(t+\tau)\right> + c.c. \label{defG32}\\ G^{(2)}(\tau)\coloneqq\left<\mathcal{E}^*(t)\mathcal{E}(t)\mathcal{E}^*(t+\tau)\mathcal{E}(t+\tau)\right>. \label{defG2}
\end{eqnarray}
Where $\mathcal{E}_{lo}(t)=A_{lo}\exp{(-i\omega_{lo}t)}$, with $A_{lo}=E_{lo}\exp{(i\theta)}$, is a coherent local oscillator with the same deterministic mode $\omega_{lo}=\avg{\omega}$ than the field. If it has to be considered as a random process, then it should be considered as independent of the field. The normalized forms are: 
\begin{equation}
    g^{(1)}(\tau)\coloneqq\frac{\avg{\mathcal{E}^*(t)\mathcal{E}(t+\tau)}}{\avg{\abs{\mathcal{E}(t)}^2}}
    \label{defg1}
\end{equation}
\begin{equation}
        g^{(3/2)}(\tau)\coloneqq \frac{1}{2} \frac{\avg{\mathcal{E}^*(t)\mathcal{E}(t)\left[\mathcal{E}^*(t+\tau)\mathcal{E}_{lo}(t+\tau)+\mathcal{E}(t+\tau)\mathcal{E}_{lo}^*(t+\tau)\right]}}{\avg{\mathcal{E}^*(t)\mathcal{E}(t)}\abs{\avg{A(t)}}\abs{\avg{A_{lo}(t)}}},
        \label{defg32}
\end{equation}
\begin{equation}
    g^{(2)}(\tau)\coloneqq\frac{\avg{\mathcal{E}^*(t)\mathcal{E}(t)\mathcal{E}^*(t+\tau)\mathcal{E}(t+\tau)}}{\avg{\mathcal{E}^*(t)\mathcal{E}(t)}\avg{\mathcal{E}^*(t+\tau)\mathcal{E}(t+\tau)}}.
    \label{defg2}
\end{equation}
The joint moment (\ref{defG32}) depends only on the time difference $\tau$ because "$\exp{(-i\delta \omega_t t)}$" and "$\exp{(i\varphi_{t+\tau})}\exp{(-i\delta \omega_{t+\tau} (t+\tau))}$" are independent, and because $\varphi_t$ is uniformly distributed over $[0;2\pi]$. 
\\The field-field correlation function $g^{(1)}(\tau)$ has already been studied in detail in reference \cite{mandel95}. References \cite{glauber2007} and \cite{handburybook} address precisely the question of what physical information lies in  the intensity-intensity correlation function $g^{(2)}(\tau)$, respectively in quantum optics, and in astronomy.

\subsection{Classical intensity-field correlation function}\label{section33}
Let $E_\beta\exp{(i\phi)}$ be the non-zero average steady part in the complex amplitude of a field $\mathcal{E}(t)$. Let $\delta A (t)$ be the fluctuations of the complex amplitude, we assume that the moments of third order for the fluctuations $\left\{\delta A(t)\right\}$ are negligible compared to the lower orders. The intensity-field correlation function $g^{(3/2)}(\tau)$ is classically defined by (\ref{defg32}). Because of the steady part $E_\beta \exp{(i\phi)}$ in the complex amplitude of the field, both the numerator and denominator in (\ref{defg32}) are non zero. We will show that this correlation function captures the evolution of a quadrature of the field, depending on the relative phase $(\phi-\theta)$ between the local oscillator and the field. \\A $\mu$-quadrature of the field, also called a ``quadrature-phase amplitude'' \cite{kimble1990quantum} of the field, is:
\begin{equation}
    A_\mu(t):=\frac{1}{\sqrt{2}}\left[A(t)\exp{(-i\mu)}+A^*(t)\exp{(i\mu)}\right].
\end{equation} 
The capture of a quadrature evolution is conditioned on an intensity fluctuation because of the term $\mathcal{E}(t)\mathcal{E}^*(t)$ in the numerator of (\ref{defg32}). In the quantum limit, it is reduced to the detection of a photon. One can already notice that (\ref{defg32}) is independent of the amplitude of the local oscillator.
\\Let us express (\ref{defg32}) in terms of the fluctuations of the field quadratures. For a given phase $\mu$ we define $\delta A_\mu(t)$ as the fluctuations of the $\mu$-quadrature of the field. After a bit of algebra, 
\begin{equation}
g^{(3/2)}(\tau) \approx \cos{(\phi-\theta)}+\frac{ \avg{ \delta A_\phi(t) \delta A_\theta(t+\tau)}}{ E_{\beta}^2 +\avg{\abs{\delta A(t)}^2}}.
\label{g32approx1}
\end{equation}
\\When the phases are equal $(\phi=\theta)$ it becomes:
\begin{equation}
g^{(3/2)}(\tau) \approx 1+\frac{ \avg{  \delta A_\theta(t) \delta A_\theta(t+\tau)}}{ E_{\beta}^2 +\avg{\abs{\delta A(t)}^2}}.
\end{equation}
First, this demonstrates that $g^{(3/2)}(\tau)$ depends on the parameters $\theta$ and $\phi$. 
Then, when $\theta=\phi$, this shows that the intensity-field correlation function captures the fluctuations of a $\theta$-quadrature of the field. It gives access to the conditional dynamics of the quadrature of the
field, similar to the manner in which the intensity-intensity correlation
function gives the conditional dynamics of the intensity
\cite{xu2015}. The results of equation \ref{g32approx1} are consistent with the quantum mechanical expressions in  \cite{carmichael04} under the appropriate classical limit.

\newpage
\section{Derivation of the correlation functions} \label{chaoticstatesection}
References \cite{carmichael00,foster00} study the intensity-field correlation function theoretically and experimentally in cavity QED from a quantum optics perspective. However, it has not been considered yet, to our knowledge, for studies of astrophysical sources such as the emission lines of ``chaotic'' nature. The study of the correlation functions (\ref{defg1}),(\ref{defg32}), and (\ref{defg2}), in the general chaotic case, where the mean of the complex amplitude in not zero, has a physical motivation. If coherent radiation exists in astrophysics, it could be observed as Rician radiation. Indeed, the radiation emerging from stellar systems is a complex combination of emission from plasma, gas, and dust. Stellar emission lines, coherent or not, could appear along with black-body radiation (the well-known zero-mean chaotic field) \cite{johansson2007astrophysical, Tan:2017}.
\subsection{Rician chaotic field}
With a steady part $E_\beta\exp{(i\phi)}$ in its complex amplitude the field becomes:
\begin{equation}
    \mathcal{E}_{\text{Ri}}(t)=A_{\text{Ri}}(t)\exp{(-i\avg{\omega}t)},
    \label{chfield2rician}
\end{equation}
where 
\begin{equation}
    A_{\text{Ri}}(t)=E_\beta\exp{(i\phi)}+A_{\text{Ra}}(t).
\end{equation}
One can show that the modulus of $A_{\text{Ri}}(t)$ follows a Rician probability law at each time, of parameters $E_\beta$ and $E_0^2/2$, with $E_0^2$ the covariance of $A_{\text{Ri}}(t)$ in its stable state. 
It is different from the common Rayleigh field (\ref{TheFieldRA}), whose modulus $\abs{A_{\text{Ra}}(t)}$ follows a Rayleigh probability law.
\\A possible physical Rician field is the superposition of a Rayleigh field emitted by our ensemble of harmonic oscillators in Brownian motion, and of coherent radiation at the same mean frequency $\avg{\omega}$. A situation that has been studied experimentally in an astrophysical context in \cite{Tan:2017}.
\subsection{Methods of calculation}\label{sectionMethodsforcorfunctions}
The calculations are done following two methods. The first one uses a common physical approach, see \cite{Loudon:book,goodman2015statistical}. The global phase $(\varphi^\ell_t-\delta\omega^\ell_t t)$ of each harmonic oscillators, is in a very good approximation uniformly distributed over $[0;2\pi]$ in the stable state of the motion. Thus $A_{\text{Ri}}(t)$ can be seen as the limit of a Pearson random walk with fixed length step \cite{kiefer1984pearson} in the complex plane. The calculation is done using a consequence of the elastic collision mechanism of Brownian motion \cite{Loudon:book,VanKampen}
\begin{equation}
    \avg{\exp{(i(\varphi^\ell_{t+\tau}-\varphi^\ell_t))}}=\exp{(-\tau/\tau_c)},
    \label{correlationofthephases}
\end{equation}
where $\tau_c$ is the mean waiting time between two collisions.
\\The second method is the use of Isserlis'theorem \cite{Isserlis1918} (that can be applied only to standardized normal random variables). Both methods give the same results.
\subsection{Results and beyond}
We introduce the aspect ratio $s$ between the width (within some constants) of the amplitude distribution and its mean
\begin{equation}
    s\coloneqq E_0^2/E_\beta^2, \label{aspectratios}
\end{equation}
and we will use the name ``Rayleigh limit'' for the limit 
\begin{equation}
    E_\beta=0 \Leftrightarrow s\to +\infty, \label{Rayleigh limit}
\end{equation}
in which the modulus of the complex amplitude distribution is the Rayleigh distribution.
\\Table \ref{tablechaosGeneral} presents the results obtained with the field (\ref{chfield2rician}) and the Rayleigh limit (\ref{TheFieldRA}). In this limit one can show that $G^{(3/2)}(\tau)$ is zero \cite{carmichael04}. One can also understand it with the circularity property of the analytic representation of a stationary zero-mean process. By definition, moments involving a different number of conjugate terms and non-conjugate terms are zero \cite{Lacoume1997}. The denominator of $g^{(3/2)}(\tau)$ is also zero in the Rayleigh limit.  The transition between the general case (Rician field) and the Rayleigh limit is in agreement with one of the experiments conducted in \cite{Tan:2017} where laser radiation (coherent) is superimposed to a black-body radiation (chaotic field, Rayleigh limit). We also get the expressions of $g^{(3/2)}(\tau)$ and $g^{(2)}(\tau)$ in terms of $g^{(1)}(\tau)$, and particularly the so-called Siegert's relation \cite{siegert1943fluctuations} for $g^{(2)}(\tau)$. When $s\to 0$ and when $s=1$, the inequalities shown in \cite{carmichael04} between $g^{(3/2)}(0)$ and $\sqrt{g^{(2)}(0)}$ are also verified here. 
The results are confirmed by our simulation detailed in section \ref{sectionsimu}.
\\The correlation functions contain two common factors: the mean waiting time between two collisions $\tau_c$ (\ref{correlationofthephases}) and the coherence time of the Doppler mechanism $\tau_\omega=\sqrt{2}/\sigma_\omega$. They form the coherence time of the radiation through the Wiener-Khintchine theorem. 
The function $g^{(3/2)}(\tau)$ depends on the local oscillator phase $\theta$ and the mean field phase $\phi$, and if $\phi=\theta$ the function tends to one when $\tau$ goes to infinity. One can already notice the dependency of $g^{(3/2)}(\tau)$ in $\abs{g^{(1)}(\tau)}$, whereas $g^{(2)}(\tau)$ depends on the square of it.
\begin{center}
\begin{table}[h!]
\caption{\label{tablechaosGeneral}Expressions of the three first 
correlation functions for a Rician and a Rayleigh field. The aspect ratio $s$ is given by (\ref{aspectratios}).
}
    \begin{tabular}{|c|c|c|}
    
    \hline
       & {Rician field} & {Rayleigh field} \\
     \hline
       & & \\
       $g^{(1)}(\tau)$ & $\dfrac{1+s\exp{(-\tau/\tau_c-\sigma_\omega^2\tau^2/2)}}{1+s}\exp{(-i\avg{\omega}\tau)}$ & $\exp{(-\tau/\tau_c-\sigma_\omega^2\tau^2/2)}\exp{(-i\avg{\omega}\tau)}$ \\
       & & \\
       \hline
       & & \\
       $g^{(3/2)}$ &  $\cos{(\theta-\phi)}\left(\dfrac{s}{1+s}+\abs{g^{(1)}(\tau)}\right)$ &  undefined \\
       & & \\
       \hline
       & & \\
       $g^{(2)}(\tau)$ & $\abs{g^{(1)}(\tau)}^2+1-\dfrac{1}{\left(1+s\right)^2}$ & $\abs{g^{(1)}(\tau)}^2+1$ \\
       & & \\
       \hline
\end{tabular}
\end{table}
\end{center}
We also consider the situation where we cannot assume the local oscillator spectrum infinitely narrow compared to the spectrum of the field. We describe the local oscillator with a probabilistic $\omega_{lo}$. 
We also assume a stable Gaussian distribution of mean $\avg{\omega_{lo}}\neq \avg{\omega}$ and variance $\sigma_{\omega_{lo}}^2$, for $\omega_{lo}$. A priori, the field and the local oscillator are independent. We get:
\begin{equation}
    g^{(3/2)}(\tau)=\cos{(\theta-\phi+(\avg{\omega_{lo}}-\avg{\omega})\tau)}\left(\frac{s}{1+s}+\abs{g^{(1)}(\tau)}\right)\exp{\left(-\frac{\sigma_{\omega_{lo}}^2}{2}\tau^2\right)}.
    \label{g32broadbandLoandfield}
\end{equation}
The beating term in $(\avg{\omega_{lo}}-\avg{\omega})\tau$ suggests that one should tune the local oscillator to get the same central frequency than the field (homodyning). The term $\exp{\left(-\sigma_{\omega_{lo}}^2\tau^2/2\right)}$ highlights the importance of having a local oscillator with a narrow spectrum.
\newpage
\section{Monte-Carlo simulation}\label{sectionsimu}
We perform a first principles Monte-Carlo simulation in order to test our calculations of table \ref{tablechaosGeneral}. The challenge is to simulate the Brownian motion and the induced fluctuations of the field. The timescales of these two processes may be very different depending on the mean frequency of the field  and the mean collision frequency. The modelling of the Brownian motion via continuous Markov processes is not precise enough to induce the proper frequency fluctuations of the field. It is necessary to model the velocity of the harmonic oscillators in the frame of the observer by a jump Markov process. For this purpose we chose an elementary model from kinetic theory that allows analytic calculation, and that reaches the Brownian motion on the macro-timescale.
\subsection{Kinetic model to reach the Brownian motion}
We use the so-called ``one dimensional gas'' at thermal equilibrium from Chapter 4 in Ref. \cite{Gillespie1991}. 
In this model each velocity component of our harmonic oscillator is a temporally homogeneous independent jump Markov process. The collision mechanism is 3D but simplified, particles can be seen as cubes whose faces are all parallel, and collisions can occur only on the faces of the cubes. They can take place off-center, but cannot engender rotations. A collision on a given face influence the velocity in the direction perpendicular to this face only. The velocity components $(v_\alpha)_\alpha$ along the axes of the cubes are independent. The collision interaction between the harmonic oscillator and the surrounding thermalised particles, is one-dimensional and independent of the velocity value. 
\subsection{Waiting times and jump-reached velocity states distributions}
The results of the calculations are similar to those following Ref.\cite{Gillespie1991}. We start from  canonical thermal equilibrium and energy and momentum conservation, we deduce the expression of the joint probability density of waiting times and velocity values reached by jumps, conditioned to a velocity value at a given time. This quantity governs where and when the velocity will jump next, which makes the simulation algorithm rather simple. 
\\The result for the joint probability density of waiting times and velocity values reached by jumps, conditioned to a velocity value, takes the form of a product of the waiting time distribution, and the density of velocity states conditioned to a velocity value. Those two functions are analytic and depend on several micro-physical parameters. First, on the reduced mass between the mass of the harmonic oscillator $m$ and the mass $m_0$ of the surrounding thermalised particles at temperature $T$ with velocity dispersion $\sigma_v=\sqrt{k_BT/m_0}$, where $k_B$ is the Boltzmann constant. Then, on the parameter $\eta=N(\varrho+\varrho_0)^2$, homogeneous to the inverse of a length, is formed with the particle density $N$ and the radii $\varrho$, of the harmonic oscillator, and $\varrho_0$, of the particles.
\subsection{Results}
The Monte-Carlo simulation is seeded with the uniform random generator from the numpy package of the python library. The other random generators are built either by a Box-Muller transform (for normal random variables), or with rejection or inversion methods. 
\\In the limit $m\gg m_0$ (for us $m\sim 10^3m_0$) and $\varrho\gg\varrho_0$ (for us $\varrho\sim 10^2\varrho_0$) one can show \cite{Gillespie1991} that the characterizing functions of the process associated to a velocity component $\alpha$ tend to the ones of an Ornstein-Uhlenbeck process, of coherence time $\tau_v=m\sqrt{\pi}/(4\sqrt{2}N\varrho^2\sqrt{m_0k_BT})$. Thus, the velocity on the timescale where it can be seen as a continuous process, tends to the Brownian motion. The continuous behavior is considered on a scale of several thousands of $\tau_c$, the mean waiting time between two collisions for the average velocity modulus $\avg{v}=2\sqrt{2k_BT}/\sqrt{\pi m}$.
\\The steady state of the velocity is reached after a few $\tau_v$. 
The stable state of the field is reached after a few $\tau_\mathcal{E}=\max{(\tau_c,\tau_\omega)}$, where $\tau_\omega=\sqrt{2}/\sigma_\omega$ is the time constant associated to the Doppler broadening $\sigma_\omega$. In our physical situation the field always reaches its stable state well before the velocity. The simulation is 3D and arbitrarily led in the atmospheric conditions of pressure at room temperature. 
The numerical precision has been set to the minimum computed waiting time, $10^5$ to $10^6$ smaller than $\tau_c$ (for us $\tau_c\sim10$ ns). 
The histograms of the three velocity components and of the Doppler frequency are consistent with theory. On the kinetic timescale the jump behavior is well represented in the velocity, and the phase jumps, and frequency changes appear clearly in the field. On the macro-timescale the Ornstein-Uhlenbeck process is reached for each velocity component. The frequency field can be tuned to increase or decrease the Doppler-Fizeau effect in front of the collision broadening. The same applies to the particles density and pressure. 
In order to have intensity fluctuations we simulated the 3D motion of three hundred independent harmonic oscillators and built the resulting field. The Rician case has been simulated by adding a coherent background to the resulting field, oscillating at the same mean frequency. 

\begin{figure}
\begin{center}
\includegraphics[width=1.0\textwidth]{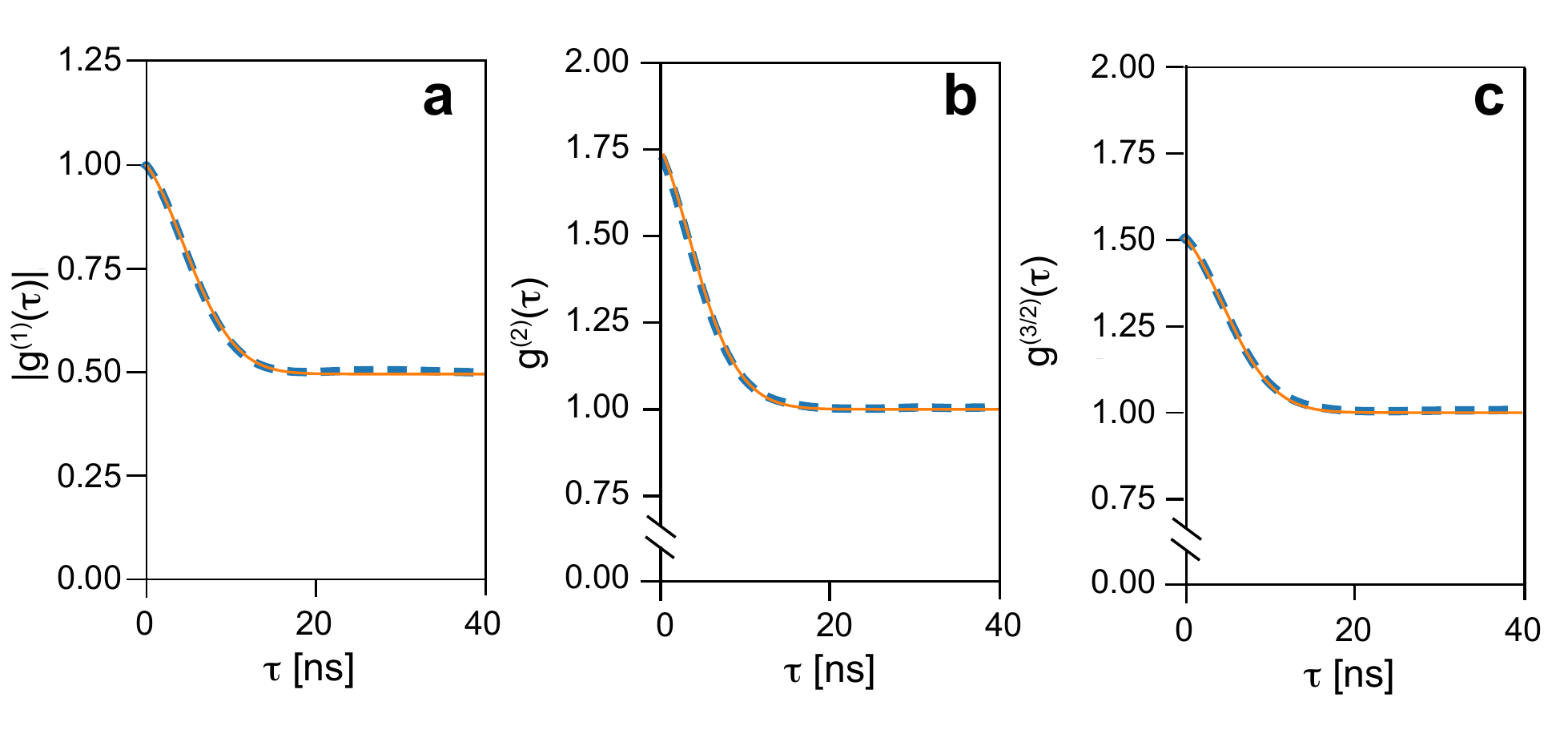}
\caption{Comparison of Monte Carlo simulations  and analytic results. The thick dashed blue curves are the simulation results for the correlation functions, the continuous orange curves are the theoretical results of table \ref{tablechaosGeneral}. In this example, the field is Rician with $s=1$ (i.e. the complex amplitude of the field dispersion is equal to its mean). The Doppler effect has slightly smaller weight than the collision broadening ($\tau_c\sim2.5\tau_{\omega}$). The number of Monte-Carlo realisations is around $5\times 10^6$. (a) is the field-field correlation function, (b) is the intensity-intensity correlation function, (c) is the intensity-field correlation function with $\phi=\theta$. }
\label{figvictory}
\end{center}
\end{figure}

The obtained correlation functions are in agreement with the calculations of table \ref{tablechaosGeneral} independent of the value of $s$ (defined (\ref{aspectratios})), and when the Doppler broadening is negligible, equal, or dominating in front of the collision broadening. We present an example of the results in figure \ref{figvictory}. In this example, the field is Rician, its coherent part and its Rayleigh part have the same weight ($s=1$). It explains why $g^{(1)}(\tau)$ does not tend to zero when $\tau\gg0$. The influence of $g^{(1)}(\tau)$ in $g^{(3/2)}(\tau)$ and $g^{(2)}(\tau)$ follows the theoretical results of table \ref{tablechaosGeneral}. Finally the classical bound derived in \cite{carmichael04}, between $g^{(3/2)}(0)$ and $\sqrt{2g^{(2)}(0)}$ is verified here for the case $s=1$. 
\color{black}
\newpage
\section{Derivation of the signal to noise ratios}\label{sectionSNR}
We compare next calculations of the signal to noise ratios (SNR) of $g^{(1)}(\tau)$, $g^{(3/2)}(\tau)$, and $g^{(2)}(\tau)$, measured by our inertial observer, in a given duration $T_0$, with a linear photo-detection system limited only by shot-noise. The statistical properties of the field are linearly coded into the statistical properties of the photocurrent $i(t)$ \cite{glauber1965}. We consider only shot-noise because it has been the main limitation in the on-stars measurements done by our group \cite{Guerin:2017}. We define shot-noise as the standard deviation of a shot-noise process. We derive the SNRs for two methods of measurements, each relevant in a different intensity regimes.
\subsection{Definitions}
The impulse response function of the linear photo-detection system is $h(t)$. We consider a decaying exponential impulse response function $h(t)=\exp{(-\gamma t)/\gamma}$ with parameter $\gamma$. $i(t)$ is the photo-current, and $e$ the charge of the photo-electrons. The shot-noise is the autocovariance of $i(t)$, taken at zero delay, when $i(t)$ is a shot-noise process. An observing duration $T_0$ creates $n_s$ independent identically distributed values $i(t)$, for a mean value $\avg{i(t)}=e\nu$ the shot-noise is:
\begin{equation}
    \delta i_m(0) = e\sqrt{\frac{\gamma\nu}{2 n_s}}.
    \label{STDshotnoiseNs}
\end{equation}
The number of samples $n_s$ is proportional to the observing time $T_0$.
For $n=1,3/2,2$, the signal is the difference between the observable $g^{(n)}(\tau)$ itself minus its value at infinite $\tau$. 
\begin{equation}
    \text{signal}\left\{g^{(n)}(\tau)\right\}\coloneqq g^{(n)}(\tau)-\lim\limits_{\tau\to+\infty}g^{(n)}(\tau).
    \label{definitondusignal}
\end{equation}
\subsection{Signal to noise ratio of $g^{(1)}(\tau)$}
The measurement of $g^{(1)}(\tau)$ through photo-detection is done by measuring a photo-current in the output of a Michelson interferometer, with a tunable path difference \cite{fox2006quantum}. 
With (\ref{definitondusignal}), for both types of chaotic fields the signal is directly given by $\avg{i(t)}\mathfrak{Re}\left\{g^{(1)}\right\}*h[\tau]$, where the average photo-current produced by the chaotic radiation is $\avg{i(t)}=e\nu$.
The noise corresponds to the standard deviation (\ref{STDshotnoiseNs}) of the shot-noise process of average density $\nu=\avg{i}/e$. The corresponding SNR is given in Table \ref{SNRtableRayleigh}. It is valid whatever the density $\nu$. For astrophysical applications as direct interferometry, where the scan of $\tau$ is done manually, there is no need to take the convolution with $h(t)$ into account. Using (\ref{STDshotnoiseNs}) and $n_s=T_0\gamma$, we get similar dependence in $\nu$ and $T_0$ to the ones in \cite{tango1980iv} and \cite{Labeyrie:book}, obtained in the SNR calculation for the spatiotemporal correlation function $g^{(1)}(\rho,\tau)$ where $\rho$ is the spatial variable. The adaptation for a Rician field is given in Table \ref{SNRtableRician}.
\subsection{Signal to noise ratios of $g^{(2)}(\tau)$ and $g^{(3/2)}(\tau)$ in the continuous regime.}
\begin{figure}
\begin{center}
\includegraphics[width=0.8\textwidth]{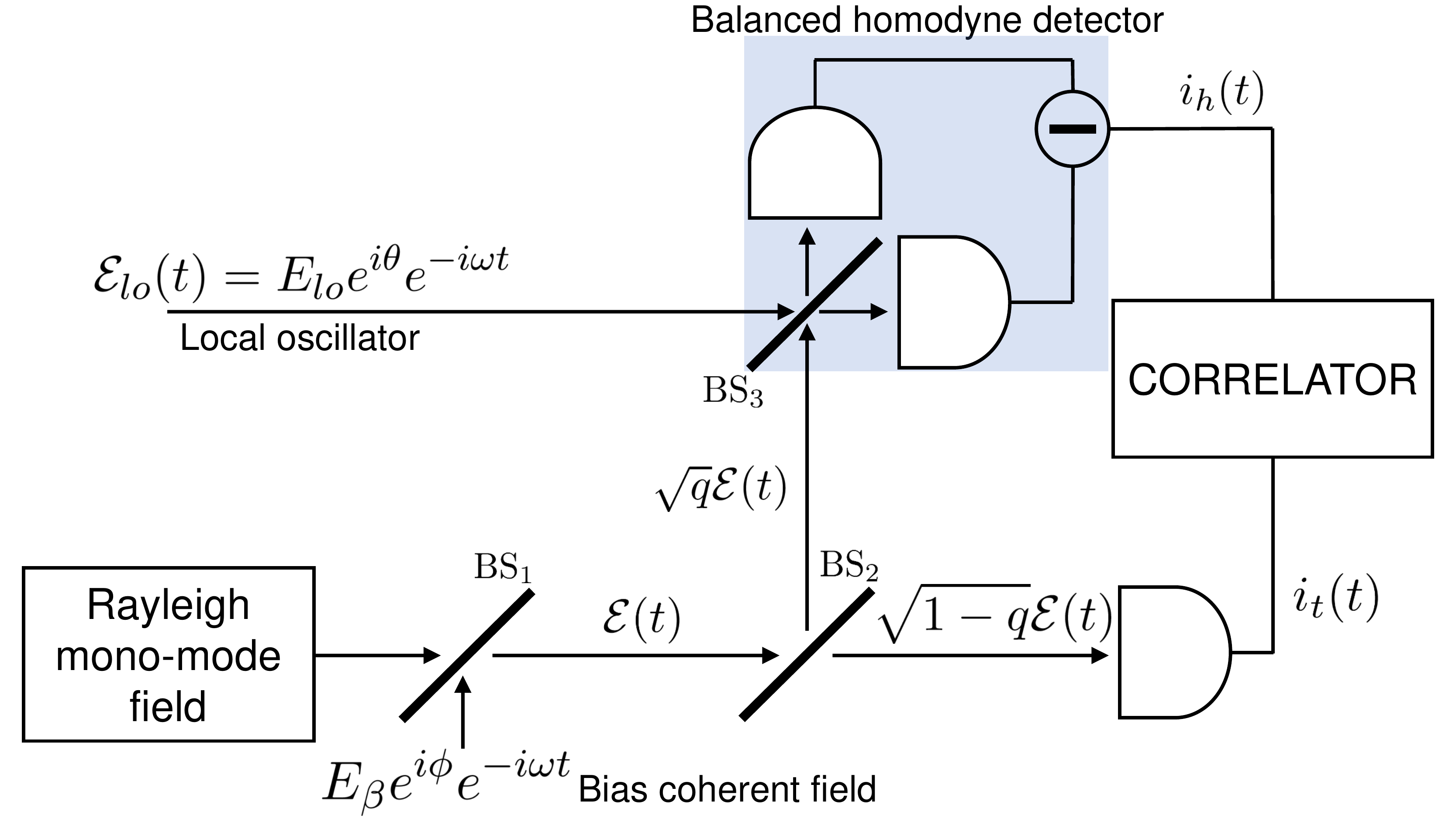}
\caption{Schematic of an optical correlator to measure the intensity-field correlation function in the example of a chaotic field with a Rayleigh amplitude. The fluctuations of the amplitude are such that $G^{(3/2)}(\tau)$ is zero. A bias coherent field is added on a first beam splitter (BS$_1$) to get a Rician amplitude. BS$_2$ sends $q$ of the intensity to a balanced homodyne detector where a local oscillator $\mathcal{E}_{\alpha}$ mixes with the field. The resulting photocurrent $i_h(t)$ is proportional to the fluctuations of a quadrature of the field. Part of the radiation intensity (with probability $(1-q)$) is directly measured on a third detector, it gives a photocurrent $i_t(t)$ that triggers the correlator. The homodyne current $i_h(t)$ averages only when the correlator is triggered.}
\label{figg32}
\end{center}
\end{figure}

For $g^{(2)}(\tau)$ and $g^{(3/2)}(\tau)$, unlike for $g^{(1)}(\tau)$, the correlation is computed by the observer and it is not directly measured from an interference phenomenon. A first method, appropriate for large densities $\nu$ called the continuous method, is the estimation of the cross-correlation via the direct-space unbiased cross-correlation estimator (the apodization function is a rectangular window). This estimator converges in mean square to $\avg{i(t)i(t+\tau)}$ when $T_0$ goes to infinity, so it converges in distribution to $\avg{i(t)i(t+\tau)}$. \\In order to evaluate the signal we use the ergodicity and the wide-sense stationarity hypothesis to express the cross-correlation as a convolution product, and deduce the signals both for $g^{(2)}(\tau)$ and $g^{(3/2)}(\tau)$. With the form of the autocorrelation of $\mathcal{E}_{\text{Ri}}(t)$ (namely $g^{(1)}(\tau)$) Slutsky's theorem \cite{slutsky1925} is verified and the autocovariance-ergodicity hypothesis is valid for $I(t)$ and so for $i(t)$.   
\\In order to evaluate the noise, we consider the variance of the cross-correlation estimator with $i(t)$ being the shot noise process. Then, because the estimator converges in distribution to $\avg{i(t)i(t+\tau)}$ for $T_0\to+\infty$, one just has to pass to the limit $1/T_0 \to 0$ in its variance to get the statistical noise on the correlation function of $i(t)$.
Notice that in the case of $g^{(3/2)}(\tau)$ the expression of the estimator involves two different currents $i_t(t)$ and $i_h(t)$, see figure \ref{figg32}. Those two noise-currents are induced by the coherent radiation of densities $\nu+\nu_\beta$ and $\nu_\alpha$ (strong local oscillator limit). We assume that the two shot-noise processes are independent and we assume that the impulse response function is the same for the two currents.
The evaluation of the noise is independent of $\nu$, and it is done by a spectral analysis in the limit of large density $\nu \to +\infty$, similarly to \cite{hanbury57}. The Fourier decomposition that we used is justified by the analysis of \cite{rice1944mathematical}.
\\Tables \ref{SNRtableRician} and \ref{SNRtableRayleigh} present the results. For $g^{(2)}(\tau)$, we get the same dependencies in $\nu$ and $T_0$ than the ones obtained in \cite{hanbury57}.  
\\In the case of $g^{(3/2)}(\tau)$, if the measurement is done with a Rayleigh amplitude field, and the offset is a bias coherent field added by the observer, then the true natural field-field correlation function $g^{(1)}(\tau)$ given in table \ref{tablechaosGeneral} for a Rayleigh field, is linked to the measured $g^{(3/2)}(\tau)$ by
\begin{equation}
    g^{(3/2)}(\tau)=\cos{(\phi-\theta)}\left(1+\frac{s}{1+s}\abs{g^{(1)}(\tau)}\right),
    \label{g32SyntheticRicianfixedomega}
\end{equation}
This is not the expression of table \ref{tablechaosGeneral}, for a Rician field. Here, the expression of $g^{(1)}(\tau)$ is for a Rayleigh field, the natural field under study. 
\subsection{Signal to noise ratios of $g^{(2)}(\tau)$ and $g^{(3/2)}(\tau)$ in the low intensity regime.}
The second method is a conditional measurement, familiar in quantum optics, it is valid only in the regime of low densities compared to the width $\gamma$ of the impulse response function $h(t)$. In this case the photo-events are rare and the continuous measurement of the previous subsection is inappropriate, the cross-correlation is estimated with the use of conditional probabilities. 
\\There are two detectors, one serves as a logical trigger, and we do not consider noise on it, the other (the balanced ``homodyne detector'', see figure \ref{figg32}) serves to record a value $i(t)$ (continuous). We do not consider any ``conditioning threshold''. The conditional measurement is the recording of a photo-event at time $\tau$, knowing that there was one at time $\tau=0$, a so-called ``click''. In this limit, a ``click'' is a non-zero value of $i(t)$. In an ideal case, on the second detector a value $i(t)$ is recorded if and only if there is $i(t)\neq 0$ on the first one. This method is legitimate only if the conditioning is not deterministic, but indeed based on the intensity fluctuations of the incoming radiation. It requires the condition $\nu \ll \gamma$. Indeed, in the case where $\nu \gg \gamma$, the ``clicks'' are going to happen almost surely, after every duration required to acquire a value $i(t)$. The conditioning becomes then deterministic (because it is periodic) and can introduce bias on the nature of the fluctuations. 
\\We use again the ergodicity and the wide-sense stationarity hypothesis to interpret the cross-correlation as a convolution product. We deduce the signals both for $g^{(2)}(\tau)$ and $g^{(3/2)}(\tau)$. The noise is directly given by (\ref{STDshotnoiseNs}) with $\nu_{lo}$ in the $g^{(3/2)}(\tau)$ case. In the $g^{(2)}(\tau)$ case, the noise is given by (\ref{STDshotnoiseNs}) with $\nu$ for a Rayleigh amplitude, and with $\nu+\nu_\beta$ for a Rician amplitude with a coherent part of density $\nu_\beta$. The results are presented in table \ref{SNRtableRician} and table \ref{SNRtableRayleigh}.
\begin{center}
\begin{table}
\caption{\label{SNRtableRician}Signal to noise ratios for a Rician field. The parameter $\nu+\nu_\beta$ is the homogeneous density of the field. 
The function $h^{(2)}(t)=h*h[t]$ is the auto-convolution of $h(t)$. For $g^{(3/2)}(\tau)$ and $g^{(2)}(\tau)$, the results are expressed in term of the field-field correlation of a Rician field given by table \ref{tablechaosGeneral}. In the upper part of the table, the results are given with precision $o\left(1/\sqrt{T_0}\right)$, and are valid regardless of the density $\nu$ of the incoming radiation. In the $g^{(1)}(\tau)$ measurement, for $\nu\gg\gamma$, the size $n_s$ of the sample is estimated from the observing time $T_0$ by $n_s=\gamma T_0$. In the lower section of the table,the results are valid if $\nu \ll \gamma$, and  the size $n_s$ of the sample is $n_s=\nu T_0$. No noise is considered on the trigger detector. The noise is the shot-noise process of the other detector.}
    \begin{tabular}{|c|c|}
    \hline
        {} & {Continuous correlation estimator} \\
     \hline
       &  \\
       $\text{SNR}\left\{g^{(1)}\right\}(\tau)$ & $\sqrt{\dfrac{2n_s}{\gamma}}\dfrac{\nu}{\sqrt{\nu+\nu_\beta}}\exp{(-\abs{t}/\tau_c-\sigma_\omega^2t^2/2)}\cos{(\avg{\omega}t)}*h(t)[\tau]$ \\
      &  \\
       \hline
       &  \\
       $\text{SNR}\left\{g^{(3/2)}\right\}(\tau)$ & $ 2\sqrt{\dfrac{T_0}{\gamma}}\cos{(\theta-\phi)}\left[\sqrt{\nu_\beta(\nu+\nu_\beta)}\abs{g^{(1)}}*h^{(2)}[\tau]-\dfrac{\nu_\beta^{3/2}}{\sqrt{\nu+\nu_\beta}}\right]$  \\
       &  \\
       \hline
       &  \\
         $\text{SNR}\left\{g^{(2)}\right\}(\tau)$ & $\sqrt{2}\sqrt{\dfrac{T_0}{\gamma}}\left[(\nu_\beta+\nu)\abs{g^{(1)}}^2*h^{(2)}[\tau]-\dfrac{\nu_\beta^2}{\nu+\nu_\beta}\right]$ \\
       &  \\
       \hline
        {} & {Conditional measure ($\nu \ll \gamma$)} \\
     \hline
       &  \\
       $\text{SNR}\left\{g^{(1)}\right\}(\tau)$ & Not available \\
       &  \\
       \hline
       &  \\
       $\text{SNR}\left\{g^{(3/2)}\right\}(\tau)$ & $2\sqrt{\dfrac{n_s}{\gamma}}\cos{(\theta-\phi)}\left[\sqrt{\nu_\beta}\abs{g^{(1)}}*h[\tau]-\dfrac{\nu_\beta^{3/2}}{\nu+\nu_\beta}\right]$  \\
       &  \\
       \hline
       &  \\
         $\text{SNR}\left\{g^{(2)}\right\}(\tau)$ & $\sqrt{\dfrac{2n_s}{\gamma}}\left[\sqrt{\nu+\nu_\beta}\abs{g^{(1)}}^2*h[\tau]-\dfrac{\nu_\beta^2}{(\nu+\nu_\beta)^{3/2}}\right]$ \\
       &  \\
       \hline
    \end{tabular}
\end{table}
\end{center}
\begin{center}
\begin{table}
\caption{\label{SNRtableRayleigh}Signal to noise ratios for a Rayleigh field. We assume that the observer adds a coherent offset of density $\nu_\beta$ in order to measure $g^{(3/2)}(\tau)$. For $g^{(3/2)}(\tau)$ and $g^{(2)}(\tau)$, the results are expressed in term of the field-field correlation function of a Rayleigh field, given by table \ref{tablechaosGeneral}. Same remarks as for table \ref{SNRtableRician}. Notice that $g^{(3/2)}(\tau)$ is optimized when $\nu \ll \nu_\beta \Leftrightarrow s \ll 1$ in the continuous method, but that $\nu_\beta$ must stay small compared to the local oscillator density $\nu_\alpha$ (otherwise this is beyond our hypothesis, and one has to take it into account in the shot noise of the homodyne current $i_h(t)$). In the conditional method $g^{(3/2)}(\tau)$ is optimized when $\nu_\beta = \nu\Leftrightarrow s=1$.}
    \begin{tabular}{|c|c|}
    \hline
        {} & {Continuous correlation estimator}  \\
     \hline
       &  \\
       $\text{SNR}\left\{g^{(1)}\right\}(\tau)$ & $\sqrt{\dfrac{2n_s}{\gamma}}\sqrt{\nu}\exp{(-\abs{t}/\tau_c-\sigma_\omega^2t^2/2)}\cos{(\avg{\omega}t)}*h(t)[\tau]$ \\
       &  \\
       \hline
       &  \\
       $\text{SNR}\left\{g^{(3/2)}\right\}(\tau)$ & $2\sqrt{\dfrac{T_0}{\gamma}}\cos{(\theta-\phi)}\dfrac{\nu\sqrt{\nu_\beta}}{\sqrt{\nu_\beta+\nu}}\abs{g^{(1)}}*h^{(2)}[\tau]$ \\
       &  \\
       \hline
       &  \\
         $\text{SNR}\left\{g^{(2)}\right\}(\tau)$ & $\sqrt{\dfrac{2T_0}{\gamma}}\nu\abs{g^{(1)}}^2*h^{(2)}[\tau]$  \\
       & \\
       \hline
        {} & {Conditional measure ($\nu \ll \gamma$)} \\
     \hline
       &  \\
       $\text{SNR}\left\{g^{(1)}\right\}(\tau)$  & Not available\\
       &  \\
       \hline
       &  \\
       $\text{SNR}\left\{g^{(3/2)}\right\}(\tau)$  & $2\sqrt{\dfrac{n_s}{\gamma}}\cos{(\theta-\phi)}\dfrac{\nu\sqrt{\nu_\beta}}{\nu+\nu_\beta}\abs{g^{(1)}}*h[\tau]$ \\
       &  \\
       \hline
       &  \\
         $\text{SNR}\left\{g^{(2)}\right\}(\tau)$  & $\sqrt{\dfrac{2n_s}{\gamma}}\sqrt{\nu}\abs{g^{(1)}}^2*h[\tau]$\\
       &  \\
       \hline
\end{tabular}
\end{table}
\end{center}
\newpage
\subsection{Comparisons}
Tables \ref{SNRtableRician} and \ref{SNRtableRayleigh} synthesize the previous results for the two methods of measurement, and for the two types of field. The results for $g^{(3/2)}(\tau)$ are valid in the strong local oscillator limit, where only the local oscillator contributes to the noise of the homodyne current $i_h(t)$. The results for the SNRs of $g^{(2)}(\tau)$ and $g^{(3/2)}(\tau)$, in table \ref{SNRtableRician}, show a competition between positive and negative contributions, both enhanced by the presence of the coherent part.
\newpage
\section{Conclusions}
We studied, for a fixed observer, the correlation functions $g^{(1)}(\tau)$, $g^{(3/2)}(\tau)$, and $g^{(2)}(\tau)$, of the field resulting from an ensemble of harmonic oscillators in Brownian motion, superimposed to a coherent background. We named this field the Rician chaotic field due to the probability distribution of its amplitude. The new results are the expressions of those correlation functions, and they are confirmed by a Monte-Carlo simulation based on the so-called ``one-dimensional gas'' model from kinetic theory. In the limit where the coherent background vanishes, our  analytic and numerical results agree with the well-established expressions for $g^{(1)}(\tau)$ and $g^{(2)}(\tau)$. 
\\Then, we derived the signal to noise ratio for an observer measuring the correlation functions with a linear photo-detection system limited by shot-noise. We did it for methods of measurement, adapted to two different intensity regimes. \\The first method, named the continuous method, gives the known dependencies for the SNRs of $g^{(1)}(\tau)$ and $g^{(2)}(\tau)$. The first new results is the SNR of $g^{(3/2)}(\tau)$. It shows the remarkable dependence in $\abs{g^{(1)}(\tau)}$. The powers in $\abs{g^{(1)}(\tau)}$ differ by one to the ones in the SNR of $g^{(2)}(\tau)$. This opens a possibility to get an on-star measurement of a coherence time twice smaller.\\For the second method, named the conditional method, the new results are the SNRs of $g^{(3/2)}(\tau)$ and $g^{(2)}(\tau)$. The powers in $\abs{g^{(1)}(\tau)}$ differ also by one. \\The implementation of a $g^{(3/2)}(\tau)$ measurement with star light is challenging because of the potentially ill-defined mean frequency of the incoming radiation, and because the local oscillator would be in reality a laser with a given spectrum. Equation (\ref{g32broadbandLoandfield}) shows that these limitations can considerably reduce the signal. However, this is a first step to open a new technique for characterizing astrophysical emission lines of chaotic nature. The possible benefit is measuring coherence times twice smaller than with $g^{(2)}(\tau)$, with a different set of technical challenges than the astronomical measurement of $g^{(1)}(\tau)$, and with the added benefit that it may characterize the non-classical fluctuations of the field (squeezing).

\section{Acknowledgments}
L. A. Orozco would like to acknowledge the hospitality of the Institute de Physique de Nice and CNRS where this work is being carried out. A. Siciak would like to thank T. Grandou and H. J. Carmichael for fruitful discussions. This work has been supported by the UCA-JEDI project ANR-15-IDEX-01. 

\clearpage
\bibliographystyle{vancouver}
\bibliography{SynthesisShortNewversion.bbl}

\end{document}